# Towards the Solution of Power Dissipation in Electronics Systems through Thermodynamics

Pradeep Singla[1], Satyan[2]

*Deptt.Of Elect.& Comm., Hindu College of Engineering, India; pardeep51355@gmail.com*
*Deptt.Of Elect.& Comm., Sonipat Institute of Engineering & Management, India; nikvsnik7@yahoo.co.in*

**ABSRACT**

**Power loss in the electronic system is a very crucial limiting fector that can be reduced or minimized with the help of using the reversible logics " a concept came from Thermodynamics".** In this paper the authors shows the concept of reversible logics for the Electronics system. The logical and physical designing approach is given in the paper in detail. The contradiction of logical and physical reversibility with the conventional CMOS designing is also shows and the solution of that contradiction is also proposed by the authors using adiabatic logic. This Paper gives a complete and clear idea if the thermodynamical concept for the electronics industries for power reduction.

## I. INTRODUCTION

In the field of cellular hardware, computer hardware, nanotechnology, bioinformatics and CMOS technology, a hardware engineer always demands for a logical IC's which dissipates very less heat/ energy or ideally no heat. But, now days the number of transistors in a IC are increasing year by year as stated by Gorden Moore (Co- founder of Intel) in 1965 called Moore's law. So, with the effect increasing the number of transistors, the heat dissipation by the hardware wills also increases. For one bit computation, KTln2 joules of energy/heat will generates as stated by Rolf Landauer in 1961[1]. Where K is a Boltzmann's constant equal to 1.3807× $10^{-23}$ $JK^{-1}$ and T is the absolute temperature at which computation is performed. This dissipated heat directly correlated to the number of lost bits. In 1973, Bennet showd that this problem can be avoided by designing the electronics hardware by reversible logics OR this dissipated energy can be saved by using reversible logics[2].

The main idea of reversible logics comes from the thermodynamics which taught us the advantage of reversible process over irreversible process. According to the second law of thermodynamics the entropy of the reversible system remains constant and doesn't dissipate heat during the process. In this we put the system or computer hardware/ electronic hardware euivalent to the heat engine. A heat engine "consumes" an amount $Q$ of high-temperature heat energy, and produces an amount $W$ of work. Thus the heat engine's efficiency is $\eta_{h.e.}$ = $W/Q$ and a computer (*i.e.*, "computational engine") consumes an amount $E_{cons}$ of free energy, and performs $N_{ops}$ useful computational operations. Thus the computer's (energy) efficiency is thus $\eta_{E,comp}$ = $N_{ops}/E_{cons}$. So, this is how the idea came in mind of making digital hardwares reversible for no heat dissipation.

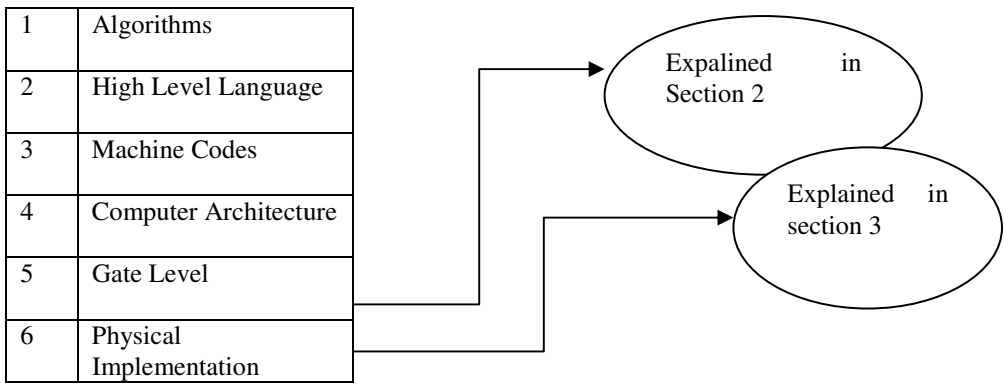

**Fig.1 Designing approach of any digital hardware**

To making the digital hardware reversible, the hardare must not only logically reversible but also must be physically reversible[3]. So, by making the digital hardwrare reversible, they will dissipates very less heat or ideally no heat or doesn't lost information i.e. they becomes adiabatic cirrcuits which recycles their energy.

There are numbers of scientist/ Researchers who are working on the different levels of the different field of energy loss reduction. The designing approach of any digital hardware is shown below.

This paper is organize in a manner that the section 2 will describe the loical reversibility at the gate level and also illustrate some reversible gates from the literatures.Section 3 will give the idea of physical implementation of the reversible hardware by adiabatic logic. This section also explain the contradiction with a statement regarding the Reversible inverter. The last section concludes the paper.

## II. LOGICAL REVERSIBLE

In case of designing any digital hardware, a hardware designer has to go for the subsystem design for the physical implementation. In this section we discuss about the gate level designing approach for the designing of reversible subsyatems. For the designing of the reversible subsyesem, the designer has to use different reversible gates which are the gates having one to one mapping between inputs and outputs or there are same number of inputs and outputs and the input can be retrievable by outputs[7-9]. These gates are diffrnent from the conventional gates because the conventional gates or irreversible gates can not regenerate its input by output But rversible can do so.The conventional inverter (NOT gate) is a reversible gate because it full fill the criteria of reversiblity i.e. having same number of input and output and the input can be regenerated by output[7-9]. But the other gates (NAND, NOR, AND, OR etc.) are not reversible.

There are numbers of reversible gates in the literature and their efficiency is based on the paramereter "Quantum Cost" which is a very important parameter in the reversible gate. Even there are numbers of parameters of the reversible gates which are describes as follows.

Constant Input (CO): The number of inputs that are kept constant ( 0 or 1) for synthesis the given functions[9].

Quantum Cost (QC): The number of reversible gates ( 1×1 or 2×2 ) to realize the circuit is known as quantum cost[9].

Garbage Output (GO): The number of outputs that are not primary is known as Garbagr outputs[9].

*I.1 Reversible Logic Gates:*

Different types of reversible gates in the literature are as follows:

Feyman Gate[5]:- Fig.2 shows the 2×2 reversible gate called Feynman gate . Feynman gate is also recognised as : controlled- not gate (CNOT). It has two inputs (A, B) and two outputs (P, Q). The outputs are defined by P=A, Q=A XOR B .This gate can be used to copy a signal. Since fan-out is not allowed in reversible logic circuits, the Feynman gate is used as the fan-out gate to copy a signal.Quantum cost of a Feyman gate is 1.

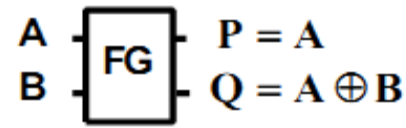

**Fig.2 Feynman Gate**

**Truth Table of Feynman gate**

| Input state | Inputs | | Outputs | | Output state |
|---|---|---|---|---|---|
| | A | B | P | Q | |
| A | 0 | 0 | 0 | 0 | a |
| B | 0 | 1 | 0 | 1 | b |
| C | 1 | 0 | 1 | 1 | d |
| D | 1 | 1 | 1 | 0 | c |

As we can see from the truth table, the inputs can be recovered if we apply output state at input side .

Toffoli gate[4]: Toffoli gate is one of the example for (3, 3) reversible gates . Fig.3 shows the Toffoli gate. Toffoli gate is also known as two controlled NOT (2-CNOT).

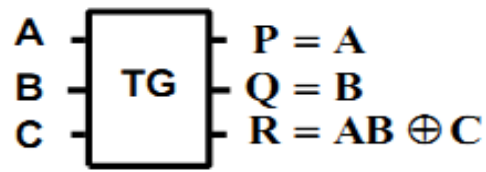

**Fig.3 Toffoli Gate**

Fredkin gate [4]: Fredkin gate is a (3, 3) reversible gate which realizes P=A, Q=A'B XOR AC and R=A'C XOR AB

where (A, B, C) is the input vector and (P, Q, R) is the output vector. Fredkin gate is also self-reversible.

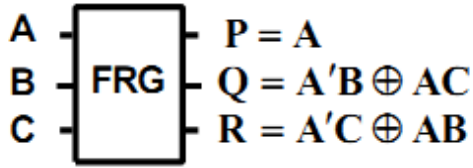

**Fig.4 Fredkin gate**

Peres gate[6]: Peres gate is another important gate which has a low quantum cost as compared to other Gate . It is shown in Fig.5.

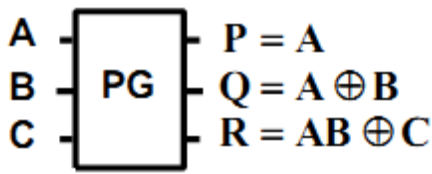

**Fig.5 Peres Gate**

New gate [9]:

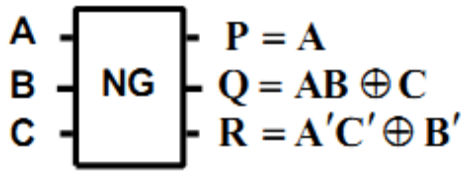

**Fig.6 New Gate**

TSG gate[9]:

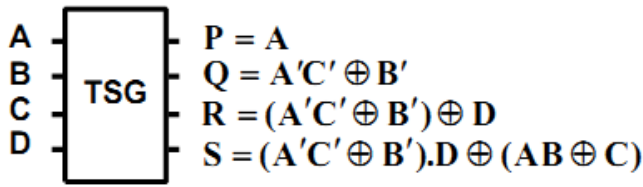

**Fig.7 TSG Gate**

**MKG gate[9]:**

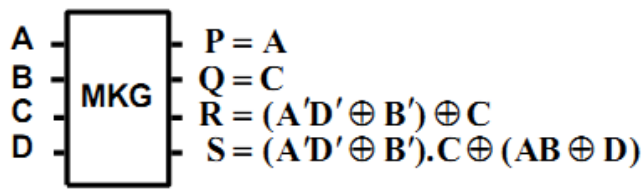

**Fig.8 MKG Gate**

So far, we have explained the different structures of the reversible gates and from the differnet figures we can see that the numbers of garbege outputs are produced by constant inputs which is a main problem of the reversible logical gates.

### III. PHYSICAL REVERSIBLE

As we have discussed in the starting section that, for making any circuit reversible, it must be reversible logically as well as physically. So, in the next we are discussing the physical reversiblity of a digital electronic hardware.

Physically reversiblity of a digital hardware can be defined in a way of system behaviour that it must be run backward also that is if a system runs backward without loss of energy that system is physically reversible[4]. In the beginning of the paper we demonstrated a leveling figure i.e. fig.1 which shows the different levels for designing any digital hardware. The gate level designing approach has been already discussed in the last section by the different strucuture or types of reversible logic gate. After that part, the physical implementation is start.

Now a days, at the transitor level design or at physical level design the designer uses CMOS due of their low power dissipation. But there are still some losses in that design when they switch that is why we force to use reversible logic. To understand the complete physical reversibility we are taking an example of CMOS invereter which is a reversible in nature because of same number of inputs and outputs and the input can be retreived by output. Now, we check whether this CMOS inverter which is a logically reversible is also physical reversible or not?

Let a CMOS inverter showing in fig. 9

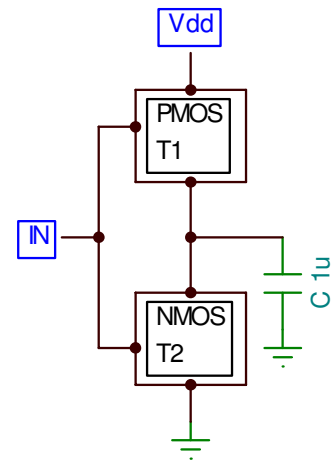

**Fig.9 conventional CMOS gate**

In this conventional CMOS logic circuit,

When Vin=0, T1 = ON , T2= OFF So, Vout = Vdd i.e. logic 1
When Vin=1, T1 = OFF , T2= ON So, Vout = 0 i.e. logic 0.

But due to the load capacitance at the ouput side, the energy stored in the capacitance at charging and discharging time is $\frac{1}{2}CVdd^2$. So, this is the energy loss/ heat produced at any time the input is changing its logical states. So, by this we can say that if we apply output voltage for retreiving the input, we can't get proper input voltage due to these energy loss. So, it makes a contradiction with the statement said in previous section that a NOT gate is a reversible gate. But, that statement was not wrong because in the conventional CMOS, R & C are the main parts of this dissipation. So, for a general form if R & C are takes in account then the energy dissipation[ 11][13] becomes

$$E = \frac{RC}{T} \times CVdd^2$$

Where T is the time for charging and discharging.
For a conventional or irreversible digital hardware, the charging time T is proportional to the RC.

So, for making it phusical reversible design we has to use a constant cuurent soure power supply uses a clock cycle is much longer than RC, i.e. trepezoidal [10-13]as shown in fig,.10

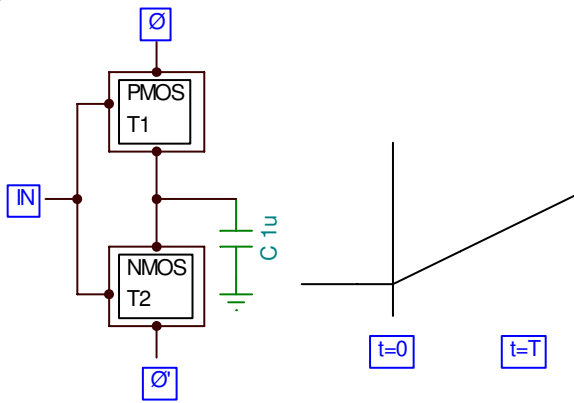

**Fig. 10 (a) Adiabatic CMOS inverter Fig.10(b) Trepezoidal waveform of Ø**

In this architecture, we use trepezoidal wave whose clock cycle is much longer than RC. The moto of using such kind of wave is to spread the charging of the gate over the whole cycle and thus reduce the energy dissipation [10-12]. This process can also be relate with the thermodynamic process of expansion of gas from a bottle with very slow movement which is a reversible process instad of expension suddenly a irreversible process. So, these kinds of circuits are adiabatic circuits which do not lost energy but a very slow process.

## IV. CONCLUSION & RESULTS

In this paper, the authors gives the complete information about the logically reversibility and physical reversibility. The interconnection of logical reversibility , physical reversibility, CMOS and adiabatic logics with thermodynamics is given in the paper in a technical manner. The authors also shows the structures of reversible gates in a logical and physical manner and also shown the merits and demerits of them. The author shows in the paper that the adiabatic reversible CMOS logic gate can be very efficient for power reduction.